\documentclass[aps,prx,twocolumn,superscriptaddress,longbibliography]{revtex4-1}

\usepackage{scalerel}
\usepackage{amssymb}
\usepackage{suffix}
\usepackage{float}
\usepackage{mathtools}
\usepackage[utf8]{inputenc}
\usepackage{booktabs}
\usepackage{cases}
\usepackage[multiple]{footmisc}
\usepackage{dcolumn}
\usepackage{color,soul}
\usepackage{rotating}
\usepackage{perpage}
\usepackage{xcolor}
\usepackage{soul}
\usepackage{tikz}
\usepackage[T1]{fontenc}
\usepackage{etoolbox}
\usepackage{graphics}
\usepackage{siunitx}
\usepackage[hidelinks]{hyperref}
\hypersetup{
	colorlinks,
	linkcolor={red},
	citecolor={blue},
	urlcolor={blue}
}
\usepackage{float}	
\usepackage{collref}
\usepackage{multirow}
\usepackage{mathtools}
\usepackage{bm}
\usepackage{url}

\usepackage{tikz}
\usepackage{tikz-3dplot}
\usepackage{changes}

\usepackage{etoolbox,lipsum}

\DeclareMathOperator{\sign}{sgn}

\begin{document} 
	\title{Probing topological phases in a perturbed Kane-Mele model via RKKY interaction: Application to monolayer jacutingaite Pt$_2$HgSe$_3$}
	
	\author{Mohsen Yarmohammadi}
	\email{mohsen.yarmohammadi@georgetown.edu}
	\address{Department of Physics, The University of Texas at Dallas, Richardson, Texas 75080, USA}
	\address{Department of Physics, Georgetown University, Washington DC 20057, USA}
	\author{Saeed Rahmanian Koshkaki}
	\address{Department of Physics, The University of Texas at Dallas, Richardson, Texas 75080, USA}
	\address{Department of Chemistry, Texas A \& M University, College Station, Texas 77842, USA}
	\author{Jamal Berakdar}
	\address{Institut f\"ur Physik, Martin-Luther Universit\"at Halle-Wittenberg, 06099 Halle/Saale, Germany}
	\author{Marin Bukov}
	\address{Max Planck Institute for the Physics of Complex Systems, N\"othnitzer Str.~38, 01187 Dresden, Germany}
	\author{Michael H. Kolodrubetz}
	\address{Department of Physics, The University of Texas at Dallas, Richardson, Texas 75080, USA}
	
	\date{\today}
	
	\begin{abstract}
		Quantum spin Hall insulators (QSHIs) are promising for spintronics, leveraging strong spin-orbit coupling for efficient spin manipulation via electrical and optical methods, with potential applications in memory storage, quantum computing, and spin-based logic. While the Kane-Mele model effectively captures the low-energy physics of these materials, the impact of perturbations driving phase transitions is less well understood. Developing approaches to describe these phases is crucial. Here, we study the noncollinear Ruderman-Kittel-Kasuya-Yosida (RKKY) interaction between two magnetic impurities in a \textit{perturbed} Kane-Mele model with strong spin-orbit coupling, relevant to monolayer jacutingaite Pt$_2$HgSe$_3$ as a prominent QSHI. By analyzing RKKY interactions, we reveal distinct, relative (rather than absolute) signatures of the different phase transitions induced by static and dynamic perturbations based on their impact on magnetic impurities. We then employ these perturbations to switch between ferromagnetic and antiferromagnetic or clockwise and counterclockwise magnetic interactions for control processes. Our results offer a practical way to track topological phases via magnetic properties.
	\end{abstract}
	
	\maketitle
	{\allowdisplaybreaks
		
		
		\section{Introduction}
		Two-dimensional (2D) quantum spin Hall insulators (QSHIs) are ultrathin materials with helical spin-polarized edge states that counterpropagate, resulting in zero net electronic conductance~\cite{doi:10.1126/science.1133734,doi:10.1126/science.1148047,doi:10.1126/science.1174736,RevModPhys.82.3045,Brune2012}. Protected by time-reversal symmetry, these states resist disorder and perturbations, allowing spin-polarized currents to travel long distances with minimal loss. The key properties of QSHIs, which rely on their topologically protected edge states, can change significantly when interacting with magnetic impurities, enabling backscattering with spin flips~\cite{PhysRevB.89.195303,PhysRevB.88.045106,PhysRevB.88.075110,PhysRevB.87.235414,PhysRevB.85.245108,PhysRevLett.106.236402,PhysRevLett.102.256803}. The Ruderman-Kittel-Kasuya-Yosida (RKKY) exchange coupling between magnetic impurities is crucial in modulating the quantum anomalous Hall effect, leading to active research on RKKY interaction for spintronic applications~\cite{PhysRevB.74.224418,PhysRevB.86.205127,PhysRevB.81.195203,Chang2015,PhysRevLett.113.137201,Awschalom2007,RevModPhys.79.1217,PhysRevB.107.054439}. Managing RKKY interactions, including ferromagnetic (FM) and antiferromagnetic (AFM) couplings and spin rotation directions (clockwise~(CW) or counterclockwise~(CCW)), is also essential, especially for magnetic data storage~\cite{SR,PhysRevB.102.075411,10.1063/5.0145144}. 
		
		Kane and Mele proposed that spin-orbit coupling (SOC) in 2D materials creates an energy band gap, resulting in a QSHI state~\cite{PhysRevLett.95.226801,PhysRevLett.95.146802}. Significant efforts are being made to explore alternative QSHIs with finite but still small SOC, such as silicene, germanene, stanene, and plumbene~\cite{PhysRevLett.108.155501,Davila_2014,Zhu2015,https://doi.org/10.1002/adma.201901017,Molle2017,Mannix2017}. Recently, jacutingaite materials (Pt$_2$AX$_3$, where A = Hg, Cd and X = S, Se) have been identified with much stronger SOC, making them promising QSHI candidates~\cite{PhysRevLett.120.117701,Marrazzo2019,PhysRevResearch.2.012063,PhysRevLett.124.106402,PhysRevB.100.235101}. The electronic behavior of jacutingaite near the Fermi level aligns with the Kane-Mele model~\cite{Kandrai2020,PhysRevLett.124.106402}. Its monolayer form exhibits unique structural properties and Fermi velocity due to the inherent strong SOC, influencing its topological and transport characteristics compared to other monolayers. 
		
		Strong SOC can alter the RKKY interaction by introducing anisotropy, affecting its strength and sign relative to the crystal axes. Exploring strong tunable SOC gaps is a cutting-edge research area that reveals various topological phases affecting the properties of quantum systems. Creating and studying new phases of matter in both equilibrium and non-equilibrium conditions can affect the inherent electronic and magnetic material properties. For instance, silicene can transition to a band insulator under external static and dynamic electric perturbations~\cite{PhysRevLett.109.055502,PhysRevLett.110.026603,PhysRevB.87.155415,PhysRevB.86.161407}. Similarly, monolayer jacutingaite Pt$_2$HgSe$_3$ can exhibit quantum anomalous Hall and quantum valley Hall effects due to magnetic exchange interactions and staggered sublattice potential~\cite{PhysRevB.106.205416}. Recent research has also investigated topological phases in Pt$_2$HgSe$_3$ under non-equilibrium conditions with high-frequency laser irradiation and periodic driving~\cite{PhysRevResearch.5.043263}.
		
		Given this background, in this work, we pose two questions when two magnetic impurities are located in a 2D QSHI with strong SOC, such as monolayer jacutingaite Pt$_2$HgSe$_3$, modeled by a perturbed Kane-Mele framework: (i) How can we identify topological phase transitions, induced by static and dynamic perturbations, via the RKKY interaction without relying on topological invariants such as the Chern number? and (ii) How can the FM/AFM or CW/CCW magnetic interactions between two impurities be switched by manipulating the host's itinerant spin and valley degrees of freedom using static and dynamic perturbations? To answer them, we use the perturbed Kane-Mele model and apply linear response theory to calculate the RKKY interaction between magnetic impurities. In exploring the static field effect, we first consider applying gate voltage on the top and bottom of the monolayer, see Fig.~\ref{f1}(a), and then consider a magnetic substrate, see Fig.~\ref{f1}(b). As for the dynamic field effect, we analyze the interaction between off-resonant circularly polarized light and monolayer jacutingaite Pt$_2$HgSe$_3$, see Fig.~\ref{f1}(c). Several studies on other systems have targeted these purposes that highlight the topic's importance~\cite{PhysRevB.109.205149,PhysRevB.99.165111,Duan2018,PhysRevB.109.184441}. However, the strong SOC in this material makes the realization of magnetic features more promising, and our procedure for detecting topological phases through RKKY interactions is both distinctive and innovative.
		
		The remainder of the paper is organized as follows. Section~\ref{s2} introduces the effective perturbed Kane-Mele Hamiltonian under static and dynamic perturbations. Section~\ref{s3} focuses on the RKKY theory. Findings are outlined in Sec.~\ref{s4}, and the paper concludes with a summary in Sec.~\ref{s5}.
		
		\section{Hamiltonian model}\label{s2}
		
		In this section, we introduce both the pristine and perturbed Kane-Mele Hamiltonian models used to describe the low-energy electronic characteristics of a 2D QSHI, as depicted in Fig.~\ref{f1} for monolayer jacutingaite Pt$_2$HgSe$_3$. The pristine Hamiltonian, which considers both spin and valley degrees of freedom and pertains to the Dirac points $\vec{K} =\tfrac{2\pi}{3a}(1,\tfrac{1}{\sqrt{3}})$ and $\vec{K}'$ $=\tfrac{2\pi}{3a}(1,-\tfrac{1}{\sqrt{3}})$ with a lattice constant $a \approx 7.4$ \AA, is generally applicable at low temperatures. It is expressed as follows~(with $\hbar = 1$)~\cite{PhysRevLett.95.226801,PhysRevLett.95.146802,PhysRevLett.110.026603,doi:10.7566/JPSJ.84.121003,PhysRevResearch.5.043263}
		\begin{equation}\label{eq_1}
		\mathcal{H}^0_{\nu , s}(\vec k) = {} v_{\rm F}(\nu k_x \sigma_x + k_y \sigma_y) + \nu s \lambda_{\rm so} \sigma_z \, ,
		\end{equation}where $\nu = +1~(-1)$ denotes the K~(K$'$) valley, and $s =+1~(-1)$ corresponds to spin up~(down). On the other hand, $v_{\rm F}$~($\approx 12.4$ eV.\AA\, for Pt$_2$HgSe$_3$) represents the Fermi velocity of electrons, while strong $\lambda_{\rm so}$ ($\approx 81.2$ meV for Pt$_2$HgSe$_3$) refers to the Kane-Mele SOC~\cite{PhysRevLett.120.117701}. In the case of monolayer Pt$_2$HgSe$_3$, the character of the Hg $s$-orbitals, which hybridize with the three nearest-neighbor Pt $d$-orbitals, significantly influences the effective above low-energy Hamiltonian. Therefore, we focus on these atoms throughout this paper, particularly when discussing the doping of magnetic impurities. The Pauli matrix vector, denoted by $\vec{\sigma}=(\sigma_x, \sigma_y, \sigma_z)$, operates on sublattice spaces with hybridized states, e.g., $|1\rangle = [|{\rm Hg},\uparrow\rangle + |{\rm Pt},\downarrow\rangle]/\sqrt{2}$ and $|2\rangle = [|{\rm Hg},\downarrow\rangle + |{\rm Pt},\uparrow\rangle]/\sqrt{2}$ in Pt$_2$HgSe$_3$~\cite{PhysRevB.105.195439}.            \begin{figure}[t]
			\centering
			\includegraphics[width=1\linewidth]{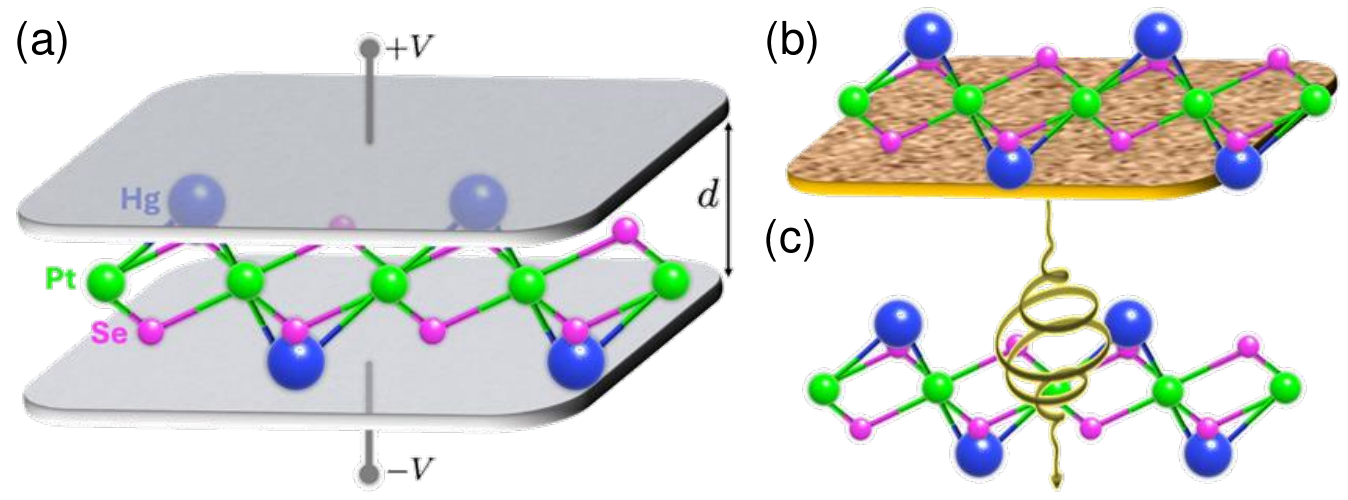}
			\caption{\textbf{Illustration of perturbed Kane-Mele monolayer jacutingaite Pt$_2$HgSe$_3$}. The structure consists of a Pt layer sandwiched between two layers of Hg and Se, arranged on a honeycomb lattice. The Hg atoms in the top and bottom layers form a buckled honeycomb lattice. In (a), the top and bottom gate electrodes induce a static electric field $E_z = V/e\,d$ with potentials $\pm V$, where $e$ represents the electron charge and $d$ denotes the vertical distance between the electrodes. In (b), the system is placed on a magnetic substrate. In (c), a continuous wave represents the dynamic electric field of light.} 
			\label{f1}
		\end{figure}
		
		To incorporate the static perturbation, we begin with the static electric field applied to a 2D QSHI. Using a combination of top and bottom gate electrodes, we generate a perpendicular electric field~\cite{PhysRevB.100.014403,Matsukura2015}. Given that the system already has an intrinsic gap due to strong SOC, this electric field mainly adjusts the SOC gap at each Dirac cone without introducing additional complexity. Extensive research has shown that this electric field can be effectively included in the Dirac Hamiltonian as a momentum-independent mass operator~\cite{PhysRevLett.109.055502,PhysRevLett.110.026603}. This results in the following Hamiltonian: $\mathcal{H}^V = {} - V \sigma_z$, with $V = e d E_z$, where $e$ represents the electron charge, $E_z$ stands for the perpendicular electric field, and $d$ indicates the vertical separation between the top and bottom electrodes, see Fig.~\ref{f1}(a). Next, we consider a static magnetic exchange field. We employ magnetic substrates, see Fig.~\ref{f1}(b), particularly those with an exchange filed $h$. This effectively modifies the SOC gap, resulting in $\mathcal{H}^h_{s} = {} s\, h\, \sigma_z$.
		
		For the dynamic perturbation, we examine the influence of circularly polarized light on the 2D QSHI Dirac cones. We introduce a time-periodic vector potential $\vec{A}(t) = A_0 [\sin(\Omega_{\rm d} t),\cos(\Omega_{\rm d} t)]$, where $A_0 = E_0/\sqrt{2} \Omega_{\rm d}$ ($E_0$ and $\Omega_{\rm d}$ represent the amplitude and frequency of the light, respectively), see Fig.~\ref{f1}(c). This choice of vector potential drives the Hamiltonian via the minimal coupling scheme, inducing transitions between eigenstates. Since the Hamiltonian exhibits periodicity in time, $\mathcal{H}^0_{\nu , s}(t) = \mathcal{H}^0_{\nu , s}(t + T)$ with period $T = 2 \pi/\Omega_{\rm d}$, the appropriate framework to address such time-periodic Hamiltonians is governed by the Floquet-Bloch theorem~\cite{GRIFONI1998229,PLATERO20041}. The Floquet Hamiltonian can be introduced and numerically diagonalized for periodically driven systems to interpret the precise photoinduced effects. However, when the electric field intensity $a = eA_0v_{\rm F}$ is relatively weak compared to the frequency $\Omega_{\rm d}$, we can approximate the Floquet Hamiltonian using a high-frequency expansion method. In this regime, the Floquet sidebands are well-separated, which prevents band crossings from creating gaps. As a result, hybridization between Floquet bands weakens, and interband electron transitions are minimized. Using a perturbative approach known as the van Vleck inverse-frequency expansion, we derive the effective Floquet Hamiltonian in powers of $\Omega_{\rm d}^{-1}$ as $\mathcal{H}_{{\rm F},\nu , s} \simeq {} \mathcal{H}^0_{\nu , s} + \frac{\left[\mathcal{H}_{-1},\mathcal{H}_{+1}\right]}{\Omega_{\rm d}}$~\cite{PhysRevB.84.235108,PhysRevB.82.235114,PhysRevB.84.235108,PhysRevLett.110.026603}, where $\mathcal{H}_{\pm 1} = {} \frac{1}{T} \int^T_0 \mathcal{H}^0_{\nu , s}(\vec{k},t) e^{\pm i \Omega_{\rm d} t} dt$. Thus, $\mathcal{H}^{\Delta_{\rm d}}_{\nu} = {} \nu \Delta_{\rm d} \sigma_z$, where a gap $\Delta_{\rm d} = {} a^2/\Omega_{\rm d}$ emerges at the Dirac cones via a two-photon process, similar to those observed before in several studies~\cite{PhysRevB.79.081406,doi:10.1063/1.3597412,PhysRevB.83.245436,PhysRevB.85.155449,PhysRevResearch.2.033228,PhysRevB.107.054439,PhysRevB.110.035307}.
		
		Hence, the total perturbed Kane-Mele Hamiltonian for a 2D QSHI, including the static and dynamic perturbations, with strong SOC reads:\begin{equation}\label{eq_5}
		\begin{aligned}
		\mathcal{H}_{\nu,s}(\vec k) = {} & v_{\rm F}(\nu k_x \sigma_x + k_y \sigma_y) + \nu s \lambda_{\rm so} \sigma_z - V \sigma_z \\ {} &+ s\, h \sigma_z + \nu \Delta_{\rm d} \sigma_z\, .
		\end{aligned}
		\end{equation} 
		Upon diagonalization of the above Hamiltonian, we obtain a gapped Dirac dispersion given by $E^{\nu , s}_{\vec k} {=} \pm\sqrt{v^2_{\rm F} k^2 + \Delta^2_{\nu,s}}$, where $+~(-)$ denotes the conduction~(valence) band. The tunable SOC gap is given by\begin{equation}\label{eq_6}
		\Delta_{\nu,s} = {} \nu s \lambda_{\rm so} - V + s \,h + \nu \Delta_{\rm d}\, .
		\end{equation}
		The static ($V$ and $h$) and dynamic ($\Delta_{\rm d}$) perturbations tune the SOC gaps for each spin and valley and then play a vital role in modulating the RKKY interaction and the magnetic interactions.

		
		\section{RKKY interaction}\label{s3}
		The indirect exchange coupling between two magnetic impurities, $\vec{S}_1$ and $\vec{S}_2$ situated at lattice sites $\vec{R}_1$ and $\vec{R}_2$ with impurity separation $\vec{R} = \vec{R}_2 - \vec{R}_1$, and mediated by the spins $\vec{s}$ of itinerant electrons in the host system, is explained by RKKY theory~\cite{10.1143/PTP.16.45,*PhysRev.106.893,*PhysRev.96.99}. In our case, these itinerant electrons are the perturbed electrons affected by both static and dynamic perturbations in a 2D QSHI, as described by Eq.~\eqref{eq_5}. Using second-order perturbation theory, the RKKY Hamiltonian is given by~\cite{Kogan,PhysRevB.83.165425,PhysRevB.84.125416}\begin{widetext}
			\begin{equation}\label{eq_7}
			\mathcal{H}^{\alpha \beta}_{\rm RKKY}(\vec{R}) = -\frac{2 J^2}{\pi} \sum_{\ell,j} S^{\ell \alpha}_1  \left({\rm Im}\int^{\mathcal{E}_{\mathrm{F}}}_{-\infty} d\mathcal{E} \mathrm{Tr}\left[\sigma_{\ell} \begin{pmatrix}
			G^{\uparrow \uparrow}_{\alpha \beta}(\mathcal{E},\vec{R}) & G^{\uparrow \downarrow}_{\alpha \beta}(\mathcal{E},\vec{R})\\\\
			G^{\downarrow \uparrow}_{\alpha \beta}(\mathcal{E},\vec{R}) & G^{\downarrow \downarrow}_{\alpha \beta}\,(\mathcal{E},\vec{R})
			\end{pmatrix}\sigma_{j}  \begin{pmatrix}
			G^{\uparrow \uparrow}_{\beta \alpha}\,(\mathcal{E},-\vec{R}) && G^{\uparrow \downarrow}_{\beta \alpha}(\mathcal{E},-\vec{R})\\\\
			G^{\downarrow \uparrow}_{\beta \alpha}(\mathcal{E},-\vec{R}) && G^{\downarrow \downarrow}_{\beta \alpha}(\mathcal{E},-\vec{R})
			\end{pmatrix} \right]\right) S^{j\beta}_2 , 
			\end{equation}\end{widetext}where $\mathcal{E}_{\rm F}$ represents the Fermi energy, $\{\alpha,\beta\}$ denote the lattice sites, and the indices $\{\ell,j\} \in \{x,y,z\}$ correspond to different spatial directions. The functions $G$ represent retarded Green's functions. 
		
		For monolayer Pt$_2$HgSe$_3$, magnetic impurities can be positioned on the same sublattices (HgHg or PtPt) or different sublattices (HgPt or PtHg). As discussed before, we focus on the Hg and Pt sublattices due to the hybridization of their orbitals which defines the effective low-energy Hamiltonian for monolayer jacutingaite, including SOC~\cite{PhysRevLett.120.117701}. Extending this to other impurity locations is straightforward, as they can be derived by adding the RKKY interaction from adjacent sites, but we omit this here to avoid complicating the calculations and results. 
		
		To calculate Green's functions, we cover the entire first Brillouin zone (FBZ) by summing over both Dirac $K$ and $K'$ points along with both spins. Therefore, we have \begin{widetext}
			\begin{subequations}\label{eq_10}
				\small \begin{align}
				&G^{s s'}_{\rm Hg Hg/Pt Pt}(\mathcal{E},\vec{R}) = \lim_{o^+ \to 0}\int \frac{d^2k}{2A_{\mathrm{FBZ}}}\, e^{i\,k\,R\,\cos\left(\varphi_k-\varphi_R\right)} \sum_{\gamma,\eta}\sum_{s'' = \pm}\left(\frac{e^{i\vec{K}\cdot \vec{R}}}{\mathcal{E}_{\nu,s''}+io^+ -\mathcal{H}_{+,s''}(\vec k)}+\frac{e^{i\vec{K}'\cdot \vec{R}}}{\mathcal{E}_{\nu,s}+io^+ -\mathcal{H}_{-,s''}(\vec k)}\right)_{\gamma,\eta} ,\\
				&G^{s s'}_{\rm Hg Pt/Pt Hg}(\mathcal{E},\vec{R}) = \lim_{o^+ \to 0}\int \frac{d^2k}{2A_{\mathrm{FBZ}}}\, e^{\pm i\,k\,R\,\cos\left(\varphi_k-\varphi_R\right)}e^{\pm i\pi/3} \sum_{\gamma,\eta}\sum_{s'' = \pm}\left(\frac{e^{\pm i\vec{K}\cdot \vec{R}}}{\mathcal{E}_{\nu,s''}+io^+ -\mathcal{H}_{+,s''}(\vec k)}-\frac{e^{\pm i\vec{K}'\cdot \vec{R}}}{\mathcal{E}_{\nu,s''}+io^+ -\mathcal{H}_{-,s''}(\vec k)}\right)_{\gamma,\eta},
				\end{align}
		\end{subequations}\end{widetext}where $A_{\mathrm{FBZ}}$ denotes the area of the entire FBZ, and $\varphi_R$ denotes the angle between $\vec R$ and $\vec K' - \vec K$, or alternatively, the angle between two magnetic impurities. Moreover, $\varphi_k$ is given by $\tan^{-1}\left(k_y/k_x\right)$ and $\{\gamma,\eta\} = \{1,2\}$ corresponds to hybridized states $|1\rangle$ and $|2\rangle$, each contributing 50\% from both spin and sublattice. 
		
		The following symmetries hold between sublattices: $G^{\uparrow \uparrow/\downarrow \downarrow}_{\rm Pt Pt}(\mathcal{E},\vec{R}) = G^{\downarrow \downarrow/\uparrow \uparrow}_{\rm Hg Hg}(\mathcal{E},\vec{R})$, $G^{\uparrow \downarrow/\downarrow \uparrow}_{\rm Pt Pt}(\mathcal{E},\vec{R}) =  G^{\downarrow \uparrow/\uparrow \downarrow}_{\rm Hg Hg}(\mathcal{E},\vec{R})$, $G^{\uparrow \uparrow/\downarrow \downarrow}_{\rm Pt Hg}(\mathcal{E},\vec{R}) = G^{\downarrow \downarrow/\uparrow \uparrow}_{\rm Hg Pt}(\mathcal{E},\vec{R})$, $G^{\uparrow \downarrow/\downarrow \uparrow}_{\rm Pt Hg}(\mathcal{E},\vec{R}) = G^{\downarrow \uparrow/\uparrow \downarrow}_{\rm Hg Pt}(\mathcal{E},\vec{R})$ and we use them in spin susceptibilities.
		
		To eliminate numerical singularities, avoid tight tolerances in real energy integration, and set the upper limit of the integral in $\mathcal{H}_{\rm RKKY}(\vec{R})$ to zero, we use the imaginary energy representation $\epsilon$, defining $\mathcal{E} + io^+ = i \epsilon$ by considering undoped phase of matter $\mathcal{E}_{\rm F} = 0$. We exclude the role of doping ($\mathcal{E}_{\rm F} \neq 0$) in this study to avoid adding complexity due to the already numerous parameters involved, but it will be analyzed in future research. The sublattice Green's functions are given in Appendix \ref{ap_1nn}. 
		
		While the RKKY interaction has been already analyzed in similar models, e.g., in silicene (with weak SOC) and graphene (without SOC) \cite{PhysRevB.81.205416,Kogan,PhysRevB.83.165425,Duan2018}, the intrinsic strong tunable SOC influences the Green's functions differently in Eq. \eqref{eq_7} through gaps $\Delta_{\zeta,s}$, which in turn, renders the RKKY interaction in perturbed 2D QSHIs more interesting compared to other materials by providing more tunability of magnetic moment of impurities.\begin{figure*}[t]
			\centering
			\includegraphics[width=0.95\linewidth]{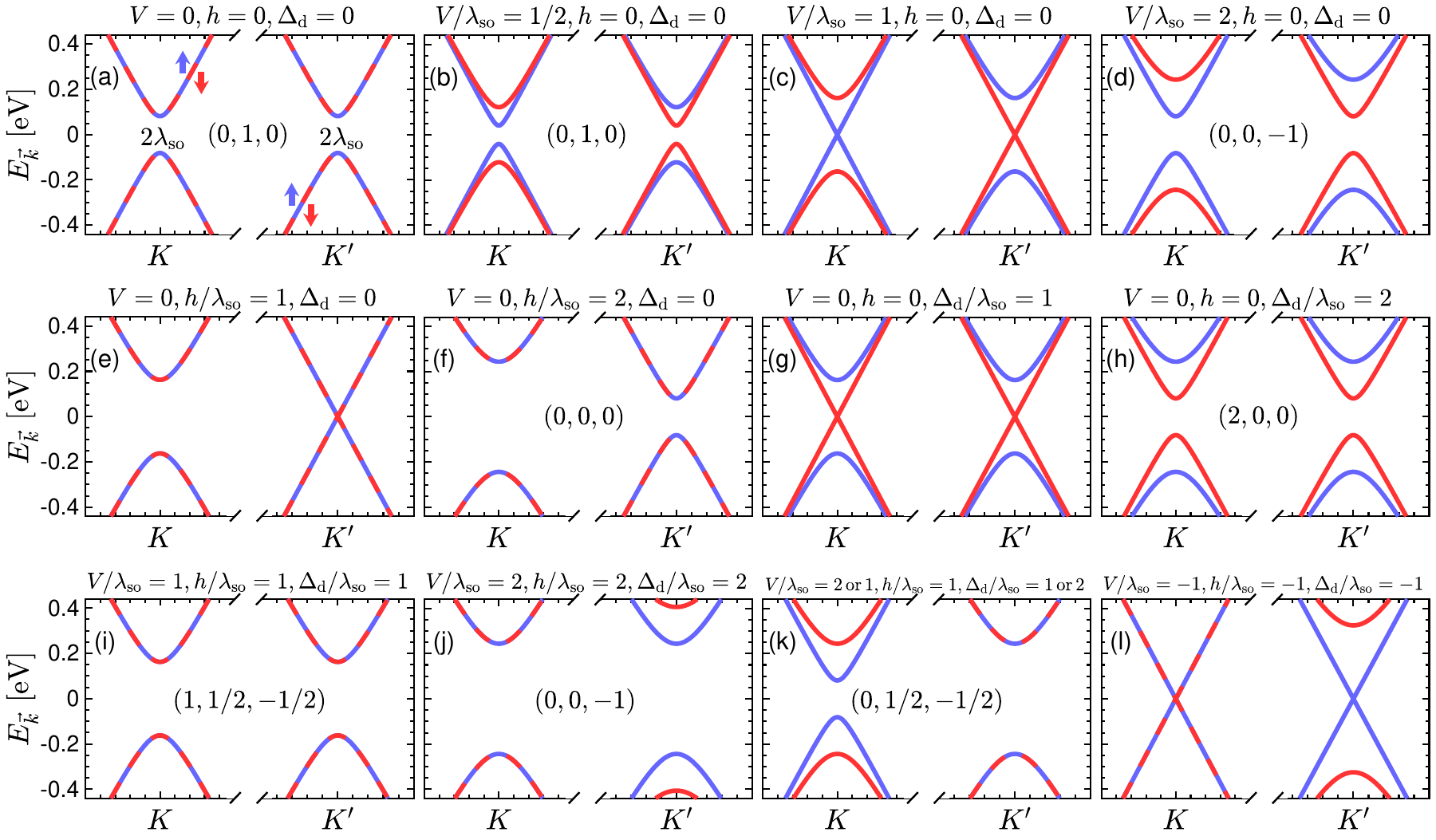}
			\caption{\textbf{The topological phase transitions~(characterized by three Chern numbers $(C,C_s,C_\nu)$) in a perturbed Kane-Mele model under static and dynamic perturbations}. In the upper row, we show electronic dispersion with (a) $V = \Delta_{\rm d} = h = 0$~(pristine QSHI with $(C,C_s,C_\nu) = (0,1,0)$), (b) $\Delta_{\rm d} = h = 0$ and $V/\lambda_{\rm so}= 1/2$~(QSHI with $(0,1,0)$), (c) $\Delta_{\rm d} = h = 0$ and $V/\lambda_{\rm so}= 1$~(spin-valley-polarized semimetal), and (d) $\Delta_{\rm d} = h = 0$ and $V/\lambda_{\rm so}= 2$~(spin-valley-polarized band insulator with $(0,0,-1)$). In the middle row, we show electronic dispersion with (e) $\Delta_{\rm d} = V = 0$ and $h/\lambda_{\rm so}= 1$~(valley-polarized single Dirac cone; a semimetal), (f) $\Delta_{\rm d} = V = 0$ and $h/\lambda_{\rm so}= 2$~(valley-polarized band insulator with $(0,0,0)$), (g) $V = h = 0$ and $\Delta_{\rm d}/\lambda_{\rm so}= 1$~(spin-polarized semimetal), and (h) $V = h = 0$ and $\Delta_{\rm d}/\lambda_{\rm so}= 2$~(quantum Hall insulator with $(2,0,0)$). In the lower row, we show electronic dispersion with (i) $\Delta_{\rm d} = V = h = \lambda_{\rm so}$~(quantum Hall insulator with $(1,1/2,-1/2)$), (j) $\Delta_{\rm d} = V = h = 2\lambda_{\rm so}$~(valley-polarized and half-spin-polarized band insulator with $(0,0,-1)$), (k) $\Delta_{\rm d} = h = \lambda_{\rm so}$ and $V/\lambda_{\rm so} = 2$ or $V = h = \lambda_{\rm so}$ and $\Delta_{\rm d}/\lambda_{\rm so} = 2$~(valley-polarized and half-spin-polarized quantum Hall insulator with $(0,1/2,-1/2)$), and (l) $\Delta_{\rm d} = V = h = -\lambda_{\rm so}$~(valley-polarized and half-spin-polarized semimetal). Overall, by adjusting the electric field, magnetic exchange field, and optical field, we can explore various phases in the system, including a QSHI with an enhanced SOC gap, a quantum Hall insulator with unique transport properties, and a semimetal with spin- and valley-dependent conduction. Following Eq.~\eqref{eq_6}, reversing $V \to -V$ causes both spin and valley to switch, changing $h \to - h$ leads to valley switching, and flipping $\Delta_{\rm d} \to - \Delta_{\rm d}$ results in spin switching. 
			} 
			\label{f2}
		\end{figure*} 
		
		Turning back to Eq.~\eqref{eq_7}, one would rewrite the RKKY Hamiltonian as $\mathcal{H}_{\rm RKKY}(\vec{R}) = - \frac{2J^2}{\pi} \text{Im} 	\int_{o^+}^{\infty} {\rm d} \epsilon i \sum_{\ell,j}S^{\ell}_1 S^{j}_2 \, \chi_{\ell j}(\epsilon,\vec{R})$, where the full expressions for the susceptibilities $\chi_{\ell j}(\epsilon,\vec{R})$ are provided in Appendix~\ref{ap1}. Accordingly, the RKKY Hamiltonian is expressed as follows:\begin{equation}\label{eq_15}
		\mathcal{H}_{\rm RKKY}(\vec{R}) = \sum_{\ell = x,y,z}J^{\rm H}_{\ell\ell}(\vec{R})S^\ell_1S^\ell_2 +\vec{J}^{\rm\,\, DM}(\vec{R}) \cdot (\vec{S}_1 \times \vec{S}_2)\,,
		\end{equation}where \begin{equation}
		J_{\ell j}(\vec{R}) = {} - \frac{2J^2}{\pi} \text{Im} 	\int_{o^+}^{\infty} {\rm d} \epsilon\, i \chi_{\ell j}(\epsilon,\vec{R})\, .
		\end{equation}
		Closed-form analytical solutions are only possible for certain specific cases already well-documented in the literature~\cite{PhysRevB.76.184430,PhysRevB.83.165425,PhysRevB.84.125416,Kogan}, so we do not repeat them here. Instead, we proceed with numerical solutions of the complicated susceptibilities. 
		
		The RKKY Hamiltonian includes both symmetric Heisenberg and asymmetric Dzyaloshinskii-Moriya~(DM) exchange interactions, which help predict the behavior of spin currents in spintronics devices. These unique spin configurations could be valuable for magnetic memory applications and might also significantly influence spin-wave behavior, potentially benefiting signal processing technologies~\cite{CAMLEY2023100605}. The Heisenberg interaction generally leads to conventional FM ($J^{\rm H}> 0$) or AFM ($J^{\rm H}< 0$) spin configurations. In contrast, the DM interaction encourages canted or helical spin structures, resulting in inversion symmetry breaking and non-collinear magnetic features with CW ($J^{\rm DM} >0$) and CCW ($J^{\rm DM} <0$) spin rotations.
		
		
		\section{Results and discussion}\label{s4}
		
		While our goal is to derive topological signatures directly from RKKY signals without relying on topological invariants like Chern numbers, we initially reference insights from the Hamiltonian band structure and Chern numbers. This ensures that the topological signatures embedded in the RKKY interactions are consistent and can be confidently and independently classified into various classes. Some of the specific topological phase transitions under static and dynamic electric perturbations presented align well with recent studies by some of us~\cite{PhysRevB.106.205416,PhysRevResearch.5.043263}, and to avoid repetition, we will not go into further details here.
		
		\subsection{Topological phase transition in a perturbed Kane-Mele model}
		
		To characterize the topological phases of a perturbed Kane-Mele system, we compute the spin-valley-resolved Chern number derived from the Berry curvature of the gapped bands, that is ultimately expressed as follows~\cite{PhysRevResearch.5.043263,PhysRevLett.110.026603}:
		\begin{equation}
		\mathcal{C}_{\nu,s} = {} \frac{\nu}{2} \sign(\nu s \lambda_{\rm so} - V + s \,h + \nu \Delta_{\rm d})\, ,
		\end{equation}
		where $\sign(...)$ is the sign function. We consider three invariants $(C,C_s,C_\nu)$: the total Chern number $C = \sum_{\nu,s} \mathcal{C}_{\nu,s}$, the spin Chern number $C_s = [\mathcal{C}_{+,+}+\mathcal{C}_{-,+}-\mathcal{C}_{+,-}-\mathcal{C}_{-,-}]/2$, and the valley Chern number $C_\nu = [\mathcal{C}_{+,+}+\mathcal{C}_{+,-}-\mathcal{C}_{-,+}-\mathcal{C}_{-,-}]/2$. The spin and valley Chern numbers can take both integer and half-integer values, which is well-known in the literature \cite{PhysRevResearch.5.043263,PhysRevLett.110.026603,Ren_2016}. These characteristics do not apply to gapless bands.
		
		Without external perturbations ($V = h = \Delta_{\rm d} = 0$), the system displays a Dirac dispersion with a $2\lambda_{\rm so}$ gap and no spin or valley polarization: a QSHI with $(C,C_s,C_\nu) = (0,1,0)$. This baseline state is highly tunable using external parameters. For instance, Figs.~\ref{f2}(b)-(d) demonstrate that increasing the electric field while keeping $\Delta_{\rm d}$ and $h$ at zero induces identical spin and valley polarization, thereby altering the material's properties. When $V$ is less than the SOC strength ($V < \lambda_{\rm so}$), the system is still in its QSHI phase with $(0,1,0)$ characters. When ($V > \lambda_{\rm so}$), the system acts as a band insulator with a full energy gap and $(0,0,-1)$. However, at the critical point $V = \lambda_{\rm so}$, the gap closes, and the system transitions into a semimetallic (SM) phase where it can conduct electricity due to the presence of a Dirac cone, indicating gapless states and higher charge mobility.
		
		When the magnetic exchange field ($h$) or optical gap ($\Delta_{\rm d}$) are varied below the SOC threshold, the phase of the system remains unchanged with $(0,1,0)$, affecting only the SOC gap. To explore different topological phases, we consider other scenarios for $h$ and $\Delta_{\rm d}$ in the following. In one case, with $h = \lambda_{\rm so}$ and $V = \Delta_{\rm d} = 0$, the system shows valley polarization without spin polarization, as seen in Fig.~\ref{f2}(e). This condition produces a single Dirac cone, indicating a SM phase. As $h$ increases, the valley polarization persists, but the Dirac cone gaps out, leading to a band-insulating phase with $(0,0,0)$, as shown in Fig.~\ref{f2}(f). In the second scenario, with both the static electric field and exchange field set to zero, varying the optical gap allows control over charge mobility and polarization, as shown in Figs.~\ref{f2}(g) and~\ref{f2}(h). Using circularly polarized light at $\Delta_{\rm d} = \lambda_{\rm so}$, spin polarization can be induced in both valleys, leading to a spin-polarized semimetal. In this phase, the spin-down direction shows a Dirac cone (SM behavior), while the other spin direction is gapped. When the optical gap exceeds the SOC strength, the system transitions to a quantum Hall insulating (QHI) phase with $(2,0,0)$.
		
		When $V = h = \Delta_{\rm d} =  \lambda_{\rm so}$, as shown in Fig.~\ref{f2}(i), the system enters a QHI phase with $(1,1/2,-1/2)$. In another scenario, when $\Delta_{\rm d} = V = h = 2\lambda_{\rm so}$ (Fig.~\ref{f2}(j)), the system exhibits a valley-polarized and half-spin-polarized band insulator with $(0,0,-1)$, where only one valley is spin-polarized and within that valley, only the spin-down state shows valley polarization. For $\Delta_{\rm d} = h = \lambda_{\rm so}$ and $V = 2\lambda_{\rm so}$ (Fig.~\ref{f2}(k)), a unique QHI phase emerges with  $(1,1/2,-1/2)$ and selective spin and valley polarization.
		
		It should be noted that reversing $V \to -V$~(reversing the top and bottom gate electrodes) results in both spin and valley switching, flipping $h \to -h$~(flipping the magnetic exchange field direction from $+z$ to $-z$) leads to only valley switching, and reversing $\Delta_{\rm d} \to -\Delta_{\rm d}$~(switching the circularly polarized light from right-handed to left-handed) causes only spin switching. All these effects lead to a different distribution of charge carriers in the system, providing more control for optimizing logic devices in experiments. For example, reversing the gate electrodes to create a static electric field in the opposite direction adjusts carrier mobility and conductivity. Flipping the AFM exchange coupling in the magnetic substrate modulates spin currents while reversing the optical gap alters the chirality of states due to the change in polarization of the induced light from right to left.
		
		A similar but distinct behavior is observed when perturbations are reversed, $\Delta_{\rm d} = V = h = -\lambda_{\rm so}$, as shown in Fig.~\ref{f2}(l). In this case, the system still exhibits valley and half-spin polarization, but the nature of the polarized states changes. Specifically, the polarized spin is gapped in one direction (indicating an insulating behavior for that spin) and SM in the other direction (indicating conducting behavior). This combination of insulating and SM behaviors within different spin channels makes the phase a valley-polarized and half-spin-polarized semimetal. The system supports SM conduction in this phase, but only for certain spin and valley configurations, which can be exploited for spintronic applications.
		
		\subsection{RKKY interaction in pristine Kane-Mele model}
		
		Before turning to the main goal of this work--investigating the topological phase outlined earlier through RKKY signals--let us first study the pristine RKKY interactions, as they play a key role in understanding the perturbed RKKY interactions later on. We note that the RKKY Hamiltonian is periodic with respect to the angle between two magnetic impurities, $\varphi_R$, as illustrated for impurities on the same sublattices in Appendix~\ref{ap2}; we proceed with $\varphi_R = \pi/3$ in the following analysis, without a specific physical justification. For simplicity, we will use the notation $\mathcal{J}(R) =\frac{\pi}{ J^2 \mathcal{A}^2} J(R)$ in the following. It is clear from the spin susceptibilities in Appendix~\ref{ap1} that each component behaves differently, and analyzing them individually in detail can lead to confusion due to the variety of possibilities. Therefore, we focus on overall in-plane~($\mathcal{J}^{\rm H}_{xx}+\mathcal{J}^{\rm H}_{yy}$ and $\mathcal{J}^{\rm DM}_{x}+\mathcal{J}^{\rm DM}_{y}$) and out-of-plane~($\mathcal{J}^{\rm H}_{zz}$ and $\mathcal{J}^{\rm DM}_{z}$) symmetric and asymmetric effects for each interaction. In particular, we avoid focusing on the magnitude of the interactions to track the sign changes associated with FM/AFM or CW/CCW orders.\begin{figure}[t]
			\centering
			\includegraphics[width=0.95\linewidth]{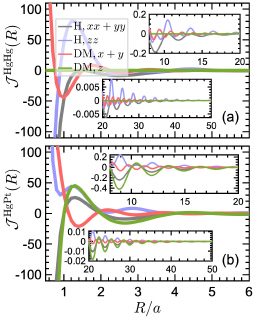}
			\caption{\textbf{RKKY interaction in pristine Kane-Mele model}. Impurity separation dependence of the RKKY interaction when magnetic impurities reside on (a) the same sublattices and (b) different sublattices at $\varphi_R = \pi/3$ and $V = h = \Delta_{\rm d} = 0$. The oscillations are strong and complex regardless of the magnetic impurities' positions on the sublattices due to the inherent strong SOC and interference effects between the K and K$'$ valleys.} 
			\label{f4}
		\end{figure}
		
		In Fig.~\ref{f4}, we illustrate the sensitivity of the pristine RKKY interaction (without static and dynamic perturbations; $V = h = \Delta_{\rm d} = 0$) to the specific lattice positions of the impurities. In both cases, the RKKY interaction strength oscillates with distance, aligning with theoretical predictions. However, the nature of these oscillations varies depending on the sublattice arrangement, underscoring the influence of crystal symmetry and electronic structure. When impurities are located on the same sublattices (HgHg), the interaction exhibits a regular oscillatory pattern for the out-of-plane Heisenberg and in-plane DM components, suggesting a tendency towards purely AFM coupling for the in-plane Heisenberg component, as shown in Fig.~\ref{f4}(a). Conversely, when impurities are situated on different sublattices (HgPt), the interaction becomes weaker and potentially more complex, characterized by larger oscillations and more pronounced FM coupling for the out-of-plane Heisenberg component, as shown in Fig.~\ref{f4}(b). Notably, we observe significantly stronger RKKY interactions in a Kane-Mele model with strong SOC compared to other models with a weak SOC or without SOC \cite{PhysRevB.81.205416,Kogan,PhysRevB.83.165425,Duan2018}.
		
		For the specific case where the $z$-component of the DM interaction vanishes when impurities reside on the same sublattices in Fig.~\ref{f4}(a), we use $\chi_{xy}$ and $\chi_{yx}$ components of the spin susceptibility in Eqs.~\eqref{eqa2d} and \eqref{eqa2e}, given by
		\begin{equation}\label{eq14a}\mathcal{J}^{\rm HgHg}_{{\rm DM},z}(R) =  {} + 8 \sin(2 k_y R_y)\, \text{Im} \int_{o^+}^{\infty} i {\rm d} \epsilon (r_- t_+ -r_+ t_-) \, ,
		\end{equation}
		where $r_{\pm}$, $t_{\pm}$, $r$, and $t$ are given in Appendix~\ref{ap1}, Eq.~\eqref{eq_A_2}. Because the terms in Eq.~\eqref{eq14a} exhibit isotropicity such that $\Delta_{+,+} = \Delta_{-,-} = \lambda_{\rm so}$ and $\Delta_{+, -} = \Delta_{-,+} = -\lambda_{\rm so}$, these contributions vanish in the pristine phase. However, this symmetry does not hold for the $x$ and $y$ components of the DM interaction, implying that there are no vanishing terms in the in-plane susceptibilities, and the system favors spin rotations exclusively in the plane.
		
		It is important to mention that a zero SOC gap produces results identical to those of graphene or a single Dirac cone in topological materials~\cite{PhysRevB.83.165425,PhysRevB.84.125416,PhysRevB.81.205416,Kogan,PhysRevB.107.054439,SR,PhysRevB.102.075411}. To avoid redundancy, we do not present it here. However, when an inherent SOC gap is present, we can derive analytical expressions for the short-distance and large-distance RKKY decay rates. In doing so, we use the following relations\begin{subequations}
			\begin{align}
			&\lim_{R \to 0} {\rm Im} \int^{\infty}_{o^+} i {\rm d} \epsilon\, \epsilon^2 K^2_n(\epsilon R/v_{\rm F}) \propto {} \frac{(2n+1) v^3_{\rm F}}{ R^3}\, , \\
			& \lim_{R \to \infty} K_n(\lambda_{\rm so} R/v_{\rm F}) \propto {}  (\lambda_{\rm so} R/v_{\rm F})^{-1/2} \,e^{- \lambda_{\rm so}R/v_{\rm F}}\label{eq_11b}\, ,
			\end{align}
		\end{subequations}resulting in $\pm R^{-3}$ decay for short-distance RKKY interactions and a combination of $\pm (\lambda_{\rm so} R)^{-1}$ decay and an exponential decay $\exp(-2\lambda_{\rm so}R/v_{\rm F})$ for large-distance interactions, in excellent agreement with Refs.~\cite{PhysRevB.83.165425,SR,Kogan,PhysRevB.107.054439}.
		
		\subsection{RKKY interaction in perturbed Kane-Mele model}
		
		Next, we explore the perturbed RKKY interaction, focusing on the distinct effects of static and dynamic electric perturbations, as shown in Fig.~\ref{f5}. The focus shifts to characterizing the topological phase transitions depicted in Fig.~\ref{f2} through the analysis of RKKY signals. Since the onset of the phase transitions occurs at small momenta around the $K$ and $K'$ points, it is important to analyze the large-distance responses to observe their fingerprint in the RKKY signals. Based on Green’s function of the non-interacting massive Dirac fermions, provided in Appendix~\ref{ap_1nn}, the expected characteristic gap-dependent length scales are $R/a \approx 2 \,v_{\rm F}$ for gating/optical driving and $R/a \approx v_{\rm F}$ for magnetic substrate effects, implying that the phases can be distinguished at scales above these thresholds. These values arise due to interference effects between the K and K$'$ valleys and different gaps induced by perturbations. Accordingly, we position the impurities relatively far apart at $R/a = 50$ for all perturbations as an alternative impurity separation to ensure that the phases begin to differentiate and continue to do so indefinitely, maintaining their distinct characteristics.
		
		We emphasize that our present signatures should be interpreted as relative rather than absolute. Specifically, while our predictions indicate, for example, an increase in FM behavior in one component and a decrease in AFM behavior in another along particular tuning axes with a fixed separation between impurities, these trends are not universally fixed. Adjustments to other parameters, such as increasing the distance between impurities, could shift the behavior--potentially leading to the discovery of a trough in the RKKY interaction instead of a peak, see Fig.~\ref{f4}. Once the shift occurs between the peak and trough, where dominant responses outweigh the vanishing ones, the sign and trend reverse without complicating the overall signatures. Thus, we present the results for peak responses in the oscillations and argue that a reverse trend occurs at the trough, which is not shown here to avoid confusion.\begin{figure*}[t]
			\centering
			\includegraphics[width=0.9\linewidth]{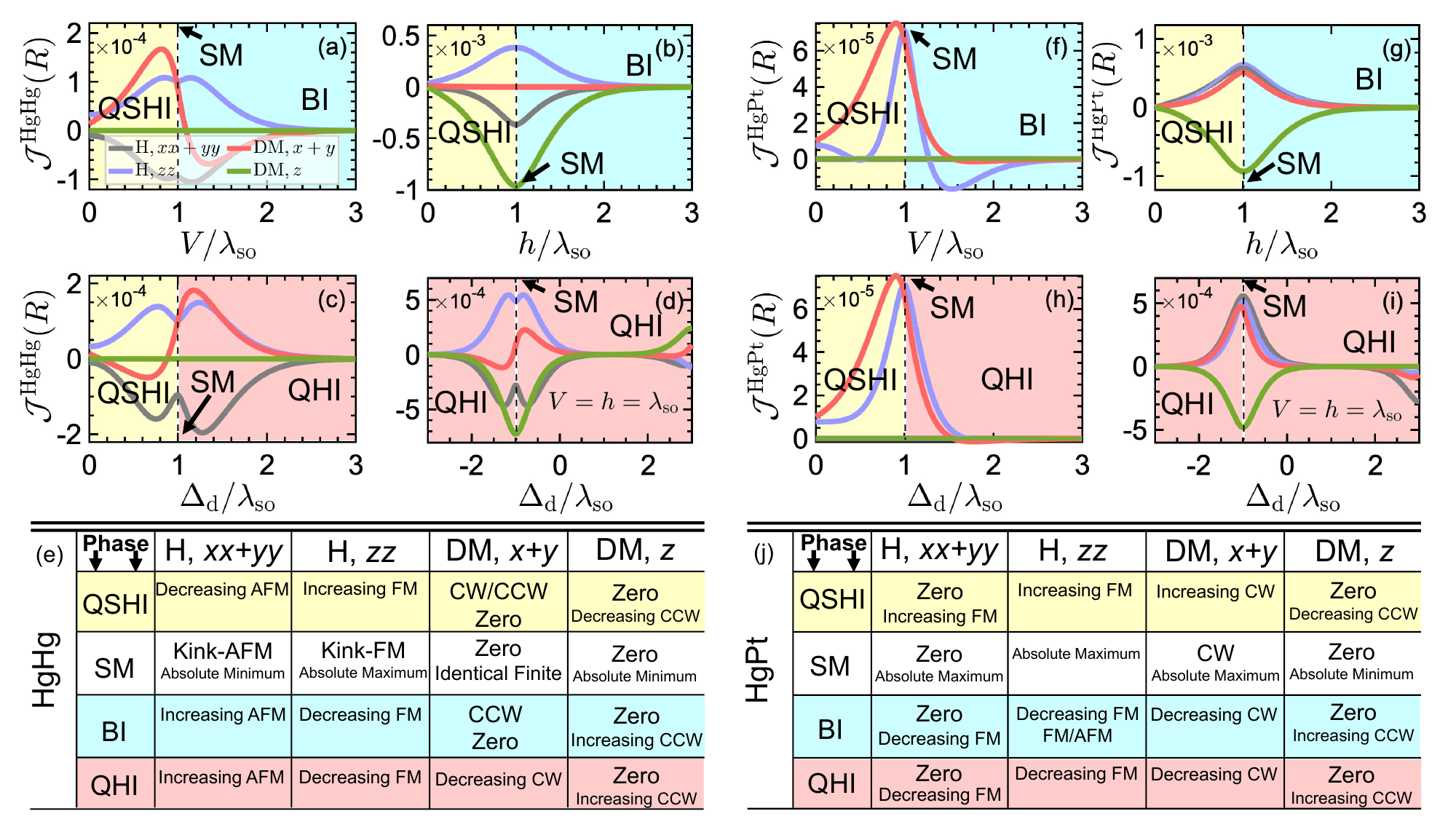}
			\caption{\textbf{RKKY interaction at large-distance between two magnetic impurities in a perturbed Kane-Mele model}. The dependence of the RKKY interactions, $\mathcal{J}(R) =\frac{A^2_{\mathrm{FBZ}}\,v^4_{\rm F}}{ J^2 \pi} J(R)$, on static and dynamic perturbations is shown for magnetic impurities located on the same HgHg sublattices ((a)-(e)), and different HgPt sublattices ((f)-(j)) at $\varphi_R = \pi/3$ and $R/a = 50$. The QSHI phase for $V = h = \Delta_{\rm d} < \lambda_{\rm so}$ transitions to the semimetallic~(SM) phase at $V = h = \Delta_{\rm d} = \lambda_{\rm so}$, and further into the band insulator (BI) or quantum Hall insulator (QHI) phases for $V = h = \Delta_{\rm d} > \lambda_{\rm so}$. The relative signatures embedded in the RKKY interactions provide a means to detect topological phases induced by static or dynamic perturbations, depending on whether magnetic impurities are placed on the same or different sublattices. All signatures for both configurations are summarized in the lower table-like panels: each row must appear simultaneously across the interaction components to accurately identify topological phases. Each second line in the box reveals a distinct signature that arises from different external stimuli, and this behavior holds for all RKKY components. The trends vary when we shift the impurity separation from a peak to a trough in Fig.~\ref{f4}, as we consider these dominant responses throughout the entire signal to consistently identify the topological phases. But they stay consistent across all RKKY components, though in reverse. For instance, if there is a decreasing AFM trend at a peak of RKKY for $R/a = 50$, we will observe an increasing FM trend at a trough with a shifted $R$. It should be noted that the trends reversing around the SM phase for the QSHI, BI, and QHI phases are a result of the SM phase itself and cannot be attributed to the other phases.} 
			\label{f5}
		\end{figure*}
		
		In the presence of a static electric field $V$, the system exhibits different \textit{isotropic} gaps: $\Delta^V_{+,+} = \Delta^V_{-,-} = \lambda_{\rm so} -V$ and $\Delta^V_{+,-} = \Delta^V_{-,+} = -\lambda_{\rm so} - V$. Similar to the pristine case, the $z$-component of the DM interaction vanishes when magnetic impurities are located on the same sublattices, as illustrated in Fig.~\ref{f5}(a). However, it also vanishes when the impurities are on different sublattices, Fig.~\ref{f5}(f). Using \eqref{eqa3d} and \eqref{eqa3e}, we have\begin{equation}\label{eq14b}
		\begin{aligned}
		\mathcal{J}^{\rm HgPt}_{{\rm DM},z}(R) = {}  & + 8  \sin(2 \varphi_R+2 \pi/3)\, \text{Im} \int_{o^+}^{\infty} i {\rm d} \epsilon\, r^2  \\ &-8\sin(2 \varphi_R-2 \pi/3) \,\text{Im}\int_{o^+}^{\infty} i {\rm d} \epsilon\, t^2 \\ &-8\sqrt{3}\cos(2 k_y R_y)\, \text{Im} \int_{o^+}^{\infty} i {\rm d} \epsilon\, r t \, .
		\end{aligned}
		\end{equation}In this case, it is due to the absence of the imaginary part of the first and third integrals in Eq.~\eqref{eq14b}, with the second term automatically being zero for $\varphi_R = \pi/3$.
		
		In the presence of a static magnetic exchange field $h$, the system experiences different \textit{anisotropic} gaps: $\Delta^h_{+,+} = -\Delta^h_{+,-} =  \lambda_{\rm so} + h$ and $\Delta^h_{-,-} = - \Delta^h_{-,+} = \lambda_{\rm so} - h$. For these gaps with such features only between spins, the $z$-component of the DM interaction does not vanish, meaning it contributes to the RKKY interactions in this case, as shown in Figs.~\ref{f5}(b) and~\ref{f5}(g); this can affect the spin alignment and magnetic properties of the system. 
		
		In the presence of a dynamic electric field $\Delta_{\rm d}$, the system exhibits different \textit{anisotropic} gaps: $\Delta^{\Delta_{\rm d}}_{+,+} = -\Delta^{\Delta_{\rm d}}_{-,+} =  \lambda_{\rm so} +\Delta_{\rm d}$ and $\Delta^{\Delta_{\rm d}}_{-,-} = - \Delta^{\Delta_{\rm d}}_{+,-} = \lambda_{\rm so} - \Delta_{\rm d}$. Due to the symmetry of these gaps between the valleys, the $z$-component of the DM interaction vanishes, as illustrated in Figs.~\ref{f5}(c) and~\ref{f5}(h). The key difference between the vanishing and non-vanishing $\mathcal{J}_{{\rm DM},z}$ in the presence of $h$ and $\Delta_{\rm d}$ can be understood from the contributions of the $r$ and $t$ terms in Eq.~\eqref{eq14b}, which primarily influence the valley degrees of freedom (see Appendix~\ref{ap1} for further details).
		
		The specific zero observed in the in-plane Heisenberg interaction, which occurs when magnetic impurities are on different sublattices and is evident under static and dynamic electric perturbations, as shown in Figs.~\ref{f5}(f) and~\ref{f5}(h), requires further clarification. It is given by~(see Appendix~\ref{ap1})\begin{equation}
		\begin{aligned}
		\mathcal{J}^{\rm HgHg}_{{\rm H},xx+yy}(R) =  {} & +8\, \text{Im} \int_{o^+}^{\infty} i {\rm d} \epsilon \,e^{-2 i \pi/3}(r_- r_+ +t_- t_+) \\{} & \hspace*{-2.25cm}- 8 \cos(2 k_y R_y)\, \text{Im} \int_{o^+}^{\infty} i {\rm d} \epsilon \,e^{-2 i \pi/3}(r_- t_+ +r_+ t_-) \, .
		\end{aligned}
		\end{equation}
		For the same reasons related to isotropic and anisotropic gaps, this term vanishes with $V$ and $\Delta_{\rm d}$. Meanwhile, the zero in-plane component of the DM interaction can be analyzed when magnetic impurities on the same sublattice are subject to a static magnetic exchange field, as illustrated in Fig.~\ref{f5}(b).
		
		After clarifying these components, we can proceed with a detailed analysis of the RKKY signals to further refine the topological phase transition characterization. The RKKY characteristics vary depending on whether the magnetic impurities are on the same or different sublattices.
		
		We stress once again that the following signatures should not be viewed as definitive or rigid evidence of precise RKKY interactions in a perturbed Kane-Mele model. Instead, they serve as a guide to the possible interaction landscape and the trends that emerge as the perturbation strength is increased or decreased.
		
		In dispersion, we previously observed that when $V < \lambda_{\rm so}$, $h < \lambda_{\rm so}$, and $\Delta_{\rm d} < \lambda_{\rm so}$, the system remains in the QSHI phase. For impurities on the same sublattices under static and dynamic electric perturbations (Figs.~\ref{f5}(a) and~\ref{f5}(c)), the AFM in-plane component of the Heisenberg interaction decreases, while the FM out-of-plane component increases, both with the same strengths ($|\tilde{J}| \approx 10^{-4}$ as a benchmark). A local minimum (for AFM) and a maximum (for FM) occur near the SM phase at $V = \Delta_{\rm d} = \lambda_{\rm so}$, where a kink is observed. These are characteristic fingerprints of the SM phase before it appears, indicating that we do not attribute such component reversals to another signature. Simultaneously, the in-plane component of the DM interaction mirrors the behavior of the out-of-plane (in-plane) Heisenberg component as a function of $V$~($\Delta_{\rm d}$). As discussed before, beyond the SM phase, static and dynamic electric perturbations drive the system into the band insulator (BI) and QHI phases. However, the BI and QHI phases can primarily be distinguished by their CW and CCW couplings in the DM interaction, which have the same strengths between magnetic impurities when $V = \Delta_{\rm d} > \lambda_{\rm so}$.
		
		While these observations are novel insights we aim to provide, the main physical argument behind them is based on the electronic density of states around the Fermi level. In the SM phase, the number of host states mediating the RKKY interaction--manifesting as a kink in the interaction--is at its maximum. Consequently, the oscillations are expected to have the largest amplitudes in the SM phase. In other phases, depending on the states involved around the Fermi level and the magnitude of the gap with respect to the inherent SOC gap, the decay behavior can also be understood through large-distance exponential decay, as described by Eq. \eqref{eq_11b}.
		
		For the static magnetic exchange field shown in Fig.~\ref{f5}(b), the situation is notably different. At $h = \lambda_{\rm so}$, we previously observed a single Dirac cone in the SM phase, characterized by an absolute minimum in both the in-plane and out-of-plane components of the Heisenberg and DM interactions. Moreover, the out-of-plane Heisenberg component reaches an absolute maximum. Beyond the SM phase, as the system transitions into the BI phase, the interactions show changes in the opposite direction.
		
		For impurities on different sublattices, the distinction between the BI and QHI phases, when $V$ and $\Delta_{\rm d}$ are above the SOC gap, is indicated by the presence (or absence) of local minima in the $z$-component of the Heisenberg interaction, as shown by the magenta lines in Figs.~\ref{f5}(f) and~\ref{f5}(h). This behavior also applies to the QSHI phase when below the SOC gap. In contrast, for $h$ in Fig.~\ref{f5}(g), the SM phase with a single Dirac cone is characterized by an absolute minimum in the $z$-component of the DM interaction.
		
		Following the same indicators for identifying topological phase transitions as previously outlined, we observe that the system enters the QHI phase when all external stimuli are 
		positive. In this phase, at equal field strengths, the RKKY interaction first vanishes and reappears, with the in-plane components showing opposite spin alignments, as illustrated in Figs.~\ref{f5}(d) and~\ref{f5}(i). A key distinction of the SM phase, under individual or combined perturbations, is the behavior of the Heisenberg interaction. With individual perturbations, the interaction shows asymmetric local minima, while combined perturbations lead to symmetric local minima. These differences appear near the critical points: $\Delta_{\rm d} = \lambda_{\rm so}$ at $V = h = 0$, as shown in Fig.~\ref{f5}(c) and $\Delta_{\rm d} = - \lambda_{\rm so}$ at $V = h = \lambda_{\rm so}$, as shown in Fig.~\ref{f5}(d). Most notably, the SM phase with combined perturbations is marked by a single absolute minimum in the $z$-component of the DM interaction when impurities reside on different sublattices, a key signature distinguishing it from other phases, as highlighted in Fig.~\ref{f5}(i).
		\begin{figure}[t]
			\centering
			\includegraphics[width=1\linewidth]{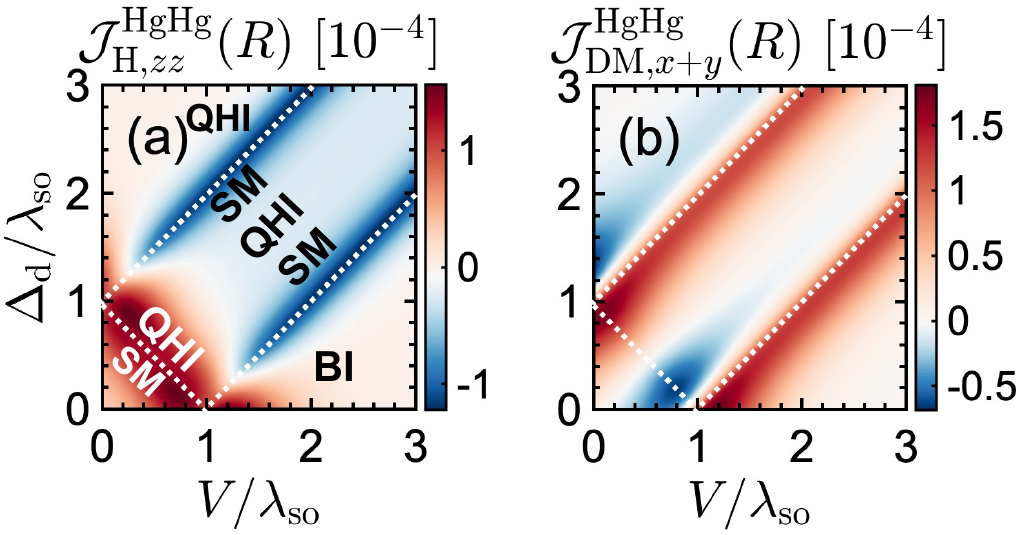}
			\caption{\textbf{RKKY interaction in perturbed Kane-Mele model}. The dependence of the RKKY interaction on the coexisting static and dynamic electric perturbations is shown for magnetic impurities located on the same sublattices at $\varphi_R = \pi/3$ and $R/a = 50$. The identification of topological phase transitions is linked to the switching between FM and AFM as well as CW and CCW magnetic interactions.} 
			\label{f6n}
		\end{figure}
		
		All the above signatures are summarized in table-like panels (e) and (j) in Fig.~\ref{f5} when magnetic impurities reside on the same and different sublattices, respectively, where the effects of perturbations are presented to simplify the classification process and clarify the distinctions between different phases.
		
		Finally, Fig.~\ref{f6n} presents the phase diagram with transitions determined by the RKKY interactions over a wide range of static and dynamic electric perturbations, with the static magnetic exchange field turned off. The absolute minima of the $z$-component of both interactions, or the reversed trends of the in-plane component of the DM interaction near the SM phase, occur along the lines defined by the equations $\Delta_{\rm d} = V \pm \lambda_{\rm so}$ or $\Delta_{\rm d} \leq \lambda_{\rm so} - V$. Additionally, for $\Delta_{\rm d} > \lambda_{\rm so} - V$, $\Delta_{\rm d} > V - \lambda_{\rm so}$, $\Delta_{\rm d} < V + \lambda_{\rm so}$, and $\Delta_{\rm d} > V + \lambda_{\rm so}$, the system is safely in the QHI phase. For $\Delta_{\rm d} < V - \lambda_{\rm so}$, the system enters the BI phase. Since there is minimal difference in the out-of-plane component of the Heisenberg interaction shown in Fig.~\ref{f6n}(a), we focus on the in-plane component in Fig.~\ref{f6n}(b), which shows the opposite spin rotations (CW and CCW). This distinction is crucial for differentiating between the BI and QHI phases. While different field configurations might yield various phase diagrams, we emphasize that the signatures identified here remain consistent across all setups. This broad applicability underscores the significance of our work, providing novel physical insights. 
		
		In addition to these findings, we observe that the magnetic ordering between impurities can be controlled by switching between FM/CW and AFM/CCW configurations with individual or combined perturbations. The positive and negative signs of the magnetic exchange couplings are indicated by the red and blue colors in Fig.~\ref{f6n}, respectively, showing the nature of the coupling. This control over magnetic ordering has potential applications in spintronic devices, where switching between FM and AFM states can influence information processing or memory storage. Precise manipulation of CW and CCW orders could also enable new functionalities in magnetic field sensors or quantum computing components.
		
		To briefly address the experimental feasibility of observing and measuring these features, we highlight that the spin orientation of each magnetic impurity can be investigated using spin-polarized scanning tunneling microscopy~\cite{Zhou2010,PhysRevLett.104.026601,Loth2010}. This capability makes the aforementioned scenarios experimentally achievable.
		
		\section{Summary}\label{s5}
		Uncovering diverse topological phases that alter the properties of quantum systems is a frontier in condensed matter research. In this work, we have identified topological signatures in a perturbed Kane-Mele 2D QSHI, applicable to a driven monolayer jacutingaite Pt$_2$HgSe$_3$ by static and dynamic perturbations, through the strong large-distance RKKY interaction~(indirect exchange coupling between two magnetic impurities in a 2D QSHI) without needing to calculate Chern numbers. Without perturbations, the system is characterized as a QSHI with a large SOC gap. Based on the information derived from topological invariants, for perturbations below the SOC gap, it remains a QSHI, while for perturbations above the SOC gap, it becomes a band or quantum Hall insulator; for specific perturbations at the SOC gap, it becomes a semimetal. 
		
		In our study of RKKY interactions, we have investigated both symmetric Heisenberg and asymmetric DM interactions to identify the above topological phases. For impurities on the same sublattices, the symmetric Heisenberg interactions reveal an absolute minimum or maximum near the semimetallic phase. When scanning impurities on different sublattices, the system typically shows semimetallic phases with either an absolute maximum or zero value in the Heisenberg interactions. Beyond the semimetallic phase, the system transitions into either a band insulator or a quantum Hall insulator phase, with the former characterized by total counterclockwise~(zero) and the latter by total clockwise~(zero) DM interactions between two impurities on the same~(different) sublattices. The observed magnetic behaviors are highly sensitive to parameter selection, potentially reversing trends when exploring different impurity separation regimes or other tunable factors. While the signatures should be understood as relative rather than absolute, the dominant RKKY responses primarily shift in sign and follow an opposite trend, without introducing additional complexity in the oscillations. Each phase offers different electronic and spin behaviors that could be useful for applications in spintronics, quantum computing, and other advanced technologies. These signatures provide practical methods to study and control topological phases through magnetic interactions, paving the way for innovative applications in various technological domains.
		
		\section*{Acknowledgments}
		M.Y. greatly thanks Fan Zhang for the helpful discussions and acknowledges the Max Planck Institute for the Physics of Complex Systems and MLU Halle-Wittenberg for their hospitality during his visit. Funded by the European Union (ERC, QuSimCtrl, 101113633). Views and opinions expressed are those of the authors and do not necessarily reflect those of the European Union or the European Research Council Executive Agency. The European Union and the granting authority are not responsible for these views. M.K. and M.Y. acknowledge support the National Science Foundation (NSF) through award numbers MPS-2228725 and DMR-1945529. J. B. acknowledges support by the DFG through SFB TRR227, B06 and project Nr. 465098690.
		
		
		\onecolumngrid
		\appendix
		\section{Sublattice Green's functions}\label{ap_1nn}
		This Appendix provides the sublattice Green's functions required to describe the spin sustainability 
		\begin{subequations}\label{eq_13}
		\small	\begin{align}
			&G^{\uparrow \uparrow/\downarrow \downarrow}_{\rm Hg Hg}(\epsilon,\vec{R}) = {}  - \mathcal{A} \sum_{s = \pm}\bigg(e^{i\vec{K}\cdot \vec{R}}\left[\left[i \epsilon\pm\Delta_{+, s}\right] K_0\left(\sqrt{ \epsilon^2 +\Delta_{+, s}^2}\,R/v_{\rm F}\right)\right] + e^{i\vec{K}'\cdot \vec{R}}\left[\left[i \epsilon\pm\Delta_{-, s}\right] K_0\left(\sqrt{ \epsilon^2 +\Delta_{-, s}^2}\,R/v_{\rm F}\right)\right]\bigg)\, ,\\
			&G^{\uparrow \downarrow/\downarrow \uparrow}_{\rm Hg Hg}(\epsilon,\vec{R}) = {}  \mp i\,\mathcal{A} \sum_{s = \pm} \bigg(e^{\pm i(\vec{K}\cdot \vec{R}+\varphi_R)}\left[\sqrt{\epsilon^2 +\Delta_{+, s}^2} \,K_1\left(\sqrt{ \epsilon^2 +\Delta_{+, s}^2}\,R/v_{\rm F}\right)\right]\notag \\ {} &\hspace{4.25cm}+ e^{\pm i (\vec{K}'\cdot \vec{R}-\varphi_R)}\left[\sqrt{\epsilon^2 +\Delta_{-, s}^2} \,K_1\left(\sqrt{ \epsilon^2+\Delta_{-, s}^2}\,R/v_{\rm F}\right)\right]\bigg)\, ,\\
			&G^{\uparrow \uparrow/\downarrow \downarrow}_{\rm Hg Pt}(\epsilon,\vec{R}) = {}  \mp i \,\mathcal{A} e^{\mp i\pi/3} \sum_{s = \pm} \bigg(e^{\pm i(\vec{K}\cdot \vec{R}+\varphi_R)}\left[\sqrt{\epsilon^2 +\Delta_{+, s}^2} \,K_1\left(\sqrt{ \epsilon^2 +\Delta_{+, s}^2}\,R/v_{\rm F}\right)\right]\notag \\ {} &\hspace{5.2cm}- e^{\pm i(\vec{K}'\cdot \vec{R}-\varphi_R)}\left[\sqrt{\epsilon^2 +\Delta_{-, s}^2} \,K_1\left(\sqrt{\epsilon^2 +\Delta_{-, s}^2}\,R/v_{\rm F}\right)\right]\bigg)\, ,\\
			&G^{\uparrow \downarrow/\downarrow \uparrow}_{\rm Hg Pt}(\epsilon,\vec{R}) = {}  - \mathcal{A} e^{- i\pi/3}\sum_{s = \pm}\bigg(e^{i\vec{K}\cdot \vec{R}}\left[\left[i\epsilon\pm\Delta_{+, s}\right] K_0\left(\sqrt{ \epsilon^2 +\Delta_{+, s}^2}\,R/v_{\rm F}\right)\right] - e^{i\vec{K}'\cdot \vec{R}}\left[\left[i\epsilon\pm\Delta_{-, s}\right] K_0\left(\sqrt{ \epsilon^2 +\Delta_{-, s}^2}\,R/v_{\rm F}\right)\right]\bigg)\, .
			\end{align}
		\end{subequations}where $\mathcal{A} = \pi/A_{\mathrm{FBZ}}\,v^2_{\rm F}$ and $K_{0,1}$ represent modified Bessel functions of the second kind. The symbols $\pm$ and $\mp$ designate different cross-talks between spins; $\{\uparrow\uparrow,\downarrow\downarrow,\uparrow\downarrow,\downarrow\uparrow\}$. 
		\section{Spin susceptibility components}\label{ap1}
		This Appendix provides the spin susceptibilities along all directions as derived from Eq.~\eqref{eq_13} and
		\begin{subequations}\label{eq1_ap_2}
			\begin{align}
			&\chi_{xx}(\epsilon,\vec{R}) =  G^{\downarrow \downarrow}(\epsilon,\vec{R})G^{\uparrow \uparrow}(\mathcal{E},-\vec{R}) + G^{\uparrow \uparrow}(\epsilon,\vec{R})G^{\downarrow \downarrow}(\mathcal{E},-\vec{R}) + G^{\downarrow \uparrow}(\epsilon,\vec{R})G^{\downarrow \uparrow}(\mathcal{E},-\vec{R}) + G^{\uparrow \downarrow}(\epsilon,\vec{R})G^{\uparrow \downarrow}(\mathcal{E},-\vec{R}) , \\
			&\chi_{yy}(\epsilon,\vec{R}) =  G^{\downarrow \downarrow}(\epsilon,\vec{R})G^{\uparrow \uparrow}(\mathcal{E},-\vec{R}) + G^{\uparrow \uparrow}(\epsilon,\vec{R})G^{\downarrow \downarrow}(\mathcal{E},-\vec{R}) - G^{\downarrow \uparrow}(\epsilon,\vec{R})G^{\downarrow \uparrow}(\mathcal{E},-\vec{R}) - G^{\uparrow \downarrow}(\epsilon,\vec{R})G^{\uparrow \downarrow}(\mathcal{E},-\vec{R}) , \\
			&\chi_{zz}(\epsilon,\vec{R}) =  G^{\uparrow \uparrow}(\epsilon,\vec{R})G^{\uparrow \uparrow}(\mathcal{E},-\vec{R}) + G^{\downarrow \downarrow}(\epsilon,\vec{R})G^{\downarrow \downarrow}(\mathcal{E},-\vec{R})  + G^{\uparrow \downarrow}(\epsilon,\vec{R})G^{\downarrow \uparrow}(\mathcal{E},-\vec{R}) + G^{\downarrow \uparrow}(\epsilon,\vec{R})G^{\uparrow \downarrow }(\mathcal{E},-\vec{R}) , \\
			&\chi_{xy}(\epsilon,\vec{R}) =  i G^{\downarrow \downarrow}(\epsilon,\vec{R})G^{\uparrow \uparrow}(\mathcal{E},-\vec{R}) - i G^{\uparrow \uparrow}(\epsilon,\vec{R})G^{\downarrow \downarrow}(\mathcal{E},-\vec{R})  + iG^{\uparrow \downarrow}(\epsilon,\vec{R})G^{\uparrow \downarrow }(\mathcal{E},-\vec{R}) -i G^{\downarrow \uparrow}(\epsilon,\vec{R})G^{\downarrow \uparrow}(\mathcal{E},-\vec{R}), \\
			&\chi_{yx}(\epsilon,\vec{R}) =  - i G^{\downarrow \downarrow}(\epsilon,\vec{R})G^{\uparrow \uparrow}(\mathcal{E},-\vec{R}) + i G^{\uparrow \uparrow}(\epsilon,\vec{R})G^{\downarrow \downarrow}(\mathcal{E},-\vec{R})  + iG^{\uparrow \downarrow}(\epsilon,\vec{R})G^{\uparrow \downarrow }(\mathcal{E},-\vec{R}) -i G^{\downarrow \uparrow}(\epsilon,\vec{R})G^{\downarrow \uparrow}(\mathcal{E},-\vec{R}), \\
			&\chi_{xz}(\epsilon,\vec{R}) =  G^{\downarrow \downarrow}(\epsilon,\vec{R})G^{\downarrow \uparrow}(\mathcal{E},-\vec{R}) + G^{\uparrow \uparrow}(\epsilon,\vec{R})G^{\uparrow \downarrow}(\mathcal{E},-\vec{R})  + G^{\downarrow \uparrow}(\epsilon,\vec{R})G^{\uparrow \uparrow }(\mathcal{E},-\vec{R}) + G^{\uparrow \downarrow}(\epsilon,\vec{R})G^{\downarrow \downarrow}(\mathcal{E},-\vec{R}), \\
			&\chi_{zx}(\epsilon,\vec{R}) =  G^{\downarrow \downarrow}(\epsilon,\vec{R})G^{\uparrow \downarrow }(\mathcal{E},-\vec{R}) + G^{\uparrow \uparrow}(\epsilon,\vec{R})G^{\downarrow \uparrow }(\mathcal{E},-\vec{R})  + G^{\downarrow \uparrow}(\epsilon,\vec{R})G^{\downarrow \downarrow}(\mathcal{E},-\vec{R}) + G^{\uparrow \downarrow}(\epsilon,\vec{R})G^{\uparrow \uparrow}(\mathcal{E},-\vec{R}), \\
			&\chi_{yz}(\epsilon,\vec{R}) = -i G^{\downarrow \downarrow}(\epsilon,\vec{R})G^{\downarrow \uparrow}(\mathcal{E},-\vec{R}) + i G^{\uparrow \uparrow}(\epsilon,\vec{R})G^{\uparrow \downarrow}(\mathcal{E},-\vec{R})  -i G^{\downarrow \uparrow}(\epsilon,\vec{R})G^{\uparrow \uparrow }(\mathcal{E},-\vec{R}) +i G^{\uparrow \downarrow}(\epsilon,\vec{R})G^{\downarrow \downarrow}(\mathcal{E},-\vec{R}), \\
			&\chi_{zy}(\epsilon,\vec{R}) = i G^{\downarrow \downarrow}(\epsilon,\vec{R})G^{\uparrow \downarrow}(\mathcal{E},-\vec{R}) - i G^{\uparrow \uparrow}(\epsilon,\vec{R})G^{\downarrow \uparrow }(\mathcal{E},-\vec{R})  +i G^{\uparrow \downarrow }(\epsilon,\vec{R})G^{\uparrow \uparrow }(\mathcal{E},-\vec{R}) -i G^{\downarrow \uparrow }(\epsilon,\vec{R})G^{\downarrow \downarrow}(\mathcal{E},-\vec{R})\, .
			\end{align}
		\end{subequations}To describe the spin susceptibilities along all directions, we simplify Eq.~\eqref{eq_13} through the following definitions
		\begin{subequations}\label{eq_A_2}
			\begin{align}
			&t_{\pm} \equiv \sum_{s = \pm}\left[\left[i\epsilon\pm\Delta_{+, s}\right] K_0\left(\sqrt{ \epsilon^2 +\Delta_{+, s}^2}\,R/v_{\rm F}\right)\right] \quad , \quad r_{\pm}\equiv\sum_{s = \pm}\left[\left[i\epsilon\pm\Delta_{-, s}\right] K_0\left(\sqrt{\epsilon^2 +\Delta_{-, s}^2}\,R/v_{\rm F}\right)\right] \,, \\
			&t\equiv\sum_{s = \pm}\left[\sqrt{ \epsilon^2 +\Delta_{+, s}^2} K_1\left(\sqrt{ \epsilon^2 +\Delta_{+, s}^2}\,R/v_{\rm F}\right)\right] \quad , \quad r\equiv\sum_{s = \pm}\left[\sqrt{ \epsilon^2 +\Delta_{-, s}^2} K_1\left(\sqrt{ \epsilon^2 +\Delta_{-, s}^2}\,R/v_{\rm F}\right)\right]\, .
			\end{align}
		\end{subequations}For magnetic impurities located on the same sublattice (HgHg), we derive
		\begin{subequations}
			\begin{align}
			\chi^{\rm HgHg}_{xx}(\epsilon,\vec{R}) ={} & + 2 \mathcal{A}^2 \left[r_- r_++t_- t_++(-2 r t+r_+ t_- +r_- t_+) \cos(2 k_y R_y)-\left(r^2+t^2\right)
			\cos(2 \varphi_R)\right]\, ,\\
			\chi^{\rm HgHg}_{yy}(\epsilon,\vec{R}) ={} &+ 2 \mathcal{A}^2 \left[r_- r_++t_- t_+ +(+2 r t+r_+ t_-+r_- t_+) \cos(2 k_y R_y)+\left(r^2+t^2\right)
			\cos(2 \varphi_R)\right]\, ,\\
			\chi^{\rm HgHg}_{zz}(\epsilon,\vec{R}) = {} &+ 2\mathcal{A}^2 \Big[\tfrac{r_-^2+r_+^2+t_-^2+t_+^2}{2}+ (r_- t_-+r_+ t_+) \cos(2 k_y R_y)+
			r^2 \cos(2 k_x R_x-2 k_y R_y)\notag \\ {} &
			+ t^2 \cos(2k_x R_x+2k_y R_y)+2 r t \cos(2 k_x
			R_x) \cos(2 \varphi_R)\Big]\, ,\\
			\chi^{\rm HgHg}_{xy}(\epsilon,\vec{R}) ={} & -2 \mathcal{A}^2 \left[(r_+ t_--r_- t_+) \sin(2 k_y R_y)+(r^2-t^2)  \sin(2 \varphi_R)\right]\, ,\label{eqa2d}\\
			\chi^{\rm HgHg}_{yx}(\epsilon,\vec{R}) ={} & +2 \mathcal{A}^2 \left[(r_+ t_--r_- t_+) \sin(2 k_y R_y)-(r^2-t^2)  \sin(2 \varphi_R)\right]\, ,\label{eqa2e}\\
			\chi^{\rm HgHg}_{xz}(\epsilon,\vec{R}) ={} &-i \mathcal{A}^2 \Big[r r_- (e^{i (2k_x R_x - 2k_y R_y+\varphi_R)} - e^{-i\varphi_R}) + r r_+ (e^{-i (2k_x R_x - 2k_y R_y-\varphi_R)} - e^{-i\varphi_R})\notag \\ {} &
			+t t_- (e^{i (2k_x R_x - 2k_y R_y-\varphi_R)} - e^{i\varphi_R}) + t t_+ (e^{-i (2k_x R_x + 2k_y R_y+\varphi_R)} - e^{i\varphi_R})\notag \\ {} &
			+r t_- (e^{i (2k_x R_x+\varphi_R)} - e^{-i(2k_y R_y+\varphi_R)}) + r t_+ (e^{-i (2k_x R_x-\varphi_R)} - e^{i(2k_y R_y-\varphi_R)})\notag \\ {} &
			+t r_- (e^{i (2k_x R_x-\varphi_R)} - e^{i(2k_y R_y+\varphi_R)}) + t r_+ (e^{-i (2k_x R_x+\varphi_R)} - e^{-i(2k_y R_y-\varphi_R)})\Big]\, ,\\
			\chi^{\rm HgHg}_{zx}(\epsilon,\vec{R}) ={} &-i \mathcal{A}^2 \Big[r r_- (e^{-i (2k_x R_x - 2k_y R_y-\varphi_R)} - e^{-i\varphi_R}) + r r_+ (e^{i (2k_x R_x - 2k_y R_y+\varphi_R)} - e^{-i\varphi_R})\notag \\ {} &
			+t t_- (e^{-i (2k_x R_x - 2k_y R_y+\varphi_R)} - e^{i\varphi_R}) + t t_+ (e^{i (2k_x R_x + 2k_y R_y-\varphi_R)} - e^{i\varphi_R})\notag \\ {} &
			+r t_- (e^{-i (2k_x R_x-\varphi_R)} - e^{i(2k_y R_y-\varphi_R)}) + r t_+ (e^{i (2k_x R_x+\varphi_R)} - e^{-i(2k_y R_y+\varphi_R)})\notag \\ {} &
			+t r_- (e^{-i (2k_x R_x+\varphi_R)} - e^{-i(2k_y R_y-\varphi_R)}) + t r_+ (e^{i (2k_x R_x-\varphi_R)} - e^{i(2k_y R_y+\varphi_R)})\Big]\, ,\\
			\chi^{\rm HgHg}_{yz}(\epsilon,\vec{R}) ={} &- \mathcal{A}^2 \Big[r r_- (e^{i (2k_x R_x - 2k_y R_y+\varphi_R)} + e^{-i\varphi_R}) + r r_+ (e^{-i (2k_x R_x - 2k_y R_y-\varphi_R)} + e^{-i\varphi_R})\notag \\ {} &
			+t t_- (e^{i (2k_x R_x + 2k_y R_y-\varphi_R)} + e^{i\varphi_R}) + t t_+ (e^{-i (2k_x R_x - 2k_y R_y+\varphi_R)} + e^{i\varphi_R})\notag \\ {} &
			+r t_- (e^{i (2k_x R_x+\varphi_R)} + e^{-i(2k_y R_y+\varphi_R)}) + r t_+ (e^{-i (2k_x R_x-\varphi_R)} + e^{i(2k_y R_y-\varphi_R)})\notag \\ {} &
			+t r_- (e^{i (2k_x R_x-\varphi_R)} + e^{i(2k_y R_y+\varphi_R)}) + t r_+ (e^{-i (2k_x R_x+\varphi_R)} + e^{-i(2k_y R_y-\varphi_R)})\Big]\, ,\\
			\chi^{\rm HgHg}_{zy}(\epsilon,\vec{R}) ={} &- \mathcal{A}^2 \Big[r r_- (e^{-i (2k_x R_x - 2k_y R_y-\varphi_R)} + e^{-i\varphi_R}) + r r_+ (e^{i (2k_x R_x - 2k_y R_y+\varphi_R)} + e^{-i\varphi_R})\notag \\ {} &
			+t t_- (e^{-i (2k_x R_x + 2k_y R_y+\varphi_R)} + e^{i\varphi_R}) + t t_+ (e^{i (2k_x R_x - 2k_y R_y-\varphi_R)} + e^{i\varphi_R})\notag \\ {} &
			+r t_- (e^{-i (2k_x R_x-\varphi_R)} + e^{i(2k_y R_y-\varphi_R)}) + r t_+ (e^{i (2k_x R_x+\varphi_R)} + e^{-i(2k_y R_y+\varphi_R)})\notag \\ {} &
			+t r_- (e^{-i (2k_x R_x+\varphi_R)} + e^{-i(2k_y R_y-\varphi_R)}) + t r_+ (e^{i (2k_x R_x-\varphi_R)} + e^{i(2k_y R_y+\varphi_R)})\Big]\, .
			\end{align}
		\end{subequations}    
		For magnetic impurities located on different sublattices (HgPt), we derive    
		\begin{subequations}
			\begin{align}
			\chi^{\rm HgPt}_{xx}(\epsilon,\vec{R}) ={} & -2 \mathcal{A}^2 \Big[r^2 \cos(2 \varphi_R + 2\pi/3) + t^2 \cos(2 \varphi_R - 2\pi/3)+ r t \cos(2
			k_y R_y) \notag \\ {} & - e^{-2 i \pi/3}\big(r_- r_+ + t_- t_+ - \cos(2 k_y R_y)\{r_- t_+ + r_+ t_-\}\big)\Big]\, ,\\
			\chi^{\rm HgPt}_{yy}(\epsilon,\vec{R}) ={} & -2 \mathcal{A}^2 \Big[r^2 \cos(2 \varphi_R + 2\pi/3) + t^2 \cos(2 \varphi_R - 2\pi/3)+ r t \cos(2
			k_y R_y) \notag \\ {} & + e^{-2 i \pi/3}\big(r_- r_+ + t_- t_+ - \cos(2 k_y R_y)\{r_- t_+ + r_+ t_-\}\big)\Big]\, ,\\
			\chi^{\rm HgPt}_{zz}(\epsilon,\vec{R}) = {} &+2\mathcal{A}^2 \Big[ r^2\cos(2 k_x R_x - 2 k_y R_y)+t^2 \cos(2 k_x R_x + 2 k_y R_y) -2 r t\cos(2 k_x R_x)\cos(2 \varphi_R) \notag \\ {} &+e^{-2 i \pi/3} \big[\tfrac{r_-^2+r_+^2+t_-^2+t_+^2}{2}- (r_- t_-+r_+ t_+) \cos(2 k_y R_y)\big]\Big]\, ,\\
			\chi^{\rm HgPt}_{xy}(\epsilon,\vec{R}) ={} & + 2 \mathcal{A}^2 \Big[r^2 \sin(2 \varphi_R + 2\pi/3) - t^2 \sin(2 \varphi_R - 2\pi/3)- \sqrt{3}r t \cos(2
			k_y R_y) \notag \\ {} &- e^{-2 i \pi/3} \sin(2 k_y R_y) \{r_+ t_- - r_- t_+\}\Big]\, ,\label{eqa3d}\\
			\chi^{\rm HgPt}_{yx}(\epsilon,\vec{R}) ={} &  + 2 \mathcal{A}^2 \Big[-r^2 \sin(2 \varphi_R + 2\pi/3) + t^2 \sin(2 \varphi_R - 2\pi/3)+ \sqrt{3}r t \cos(2
			k_y R_y) \notag \\ {} &- e^{-2 i \pi/3} \sin(2 k_y R_y) \{r_+ t_- - r_- t_+\}\Big]\, ,\label{eqa3e}\\
			\chi^{\rm HgPt}_{xz}(\epsilon,\vec{R}) ={} &i \mathcal{A}^2 e^{-2 i \pi/3} \Big[r r_- (-e^{i (2k_x R_x-2k_y R_y+\varphi_R + 2\pi/3)} + e^{-i\varphi_R}) + r r_+ (-e^{i (-2k_x R_x+2k_y R_y+\varphi_R + 2\pi/3)} + e^{-i\varphi_R})\notag \\ {} &
			+t t_- (-e^{i (2k_x R_x + 2k_y R_y-\varphi_R+ 2\pi/3)} + e^{i\varphi_R}) + t t_+ (-e^{i (-2k_x R_x - 2k_y R_y-\varphi_R+ 2\pi/3)} + e^{i\varphi_R})\notag \\ {} &
			-r t_- (e^{-i (2k_y R_y+\varphi_R)} - e^{i(2k_x R_x+\varphi_R+2\pi/3)}) - r t_+ (e^{i (2k_y R_y-\varphi_R)} - e^{-i(2k_x R_x-\varphi_R-2\pi/3)})\notag \\ {} &
			-t r_- (e^{i (2k_y R_y+\varphi_R)} - e^{i(2k_x R_x-\varphi_R+2\pi/3)}) - t r_+ (e^{-i (2k_y R_y-\varphi_R)} - e^{-i(2k_x R_x+\varphi_R-2\pi/3)})\Big]\, ,\\
			\chi^{\rm HgPt}_{zx}(\epsilon,\vec{R}) ={} &i \mathcal{A}^2 e^{-2 i \pi/3} \Big[r r_- (-e^{i (-2k_x R_x+2k_y R_y+\varphi_R + 2\pi/3)} + e^{-i\varphi_R}) + r r_+ (-e^{i (2k_x R_x-2k_y R_y+\varphi_R + 2\pi/3)} + e^{-i\varphi_R})\notag \\ {} &
			+t t_- (-e^{-i (2k_x R_x + 2k_y R_y+\varphi_R- 2\pi/3)} + e^{i\varphi_R}) + t t_+ (-e^{i (2k_x R_x + 2k_y R_y-\varphi_R+ 2\pi/3)} + e^{i\varphi_R})\notag \\ {} &
			-r t_- (e^{i (2k_y R_y-\varphi_R)} - e^{i(-2k_x R_x+\varphi_R+2\pi/3)}) - r t_+ (e^{i (-2k_y R_y-\varphi_R)} - e^{i(2k_x R_x+\varphi_R+2\pi/3)})\notag \\ {} &
			-t r_- (e^{i (-2k_y R_y+\varphi_R)} - e^{i(-2k_x R_x-\varphi_R+2\pi/3)}) - t r_+ (e^{i (2k_y R_y+\varphi_R)} - e^{i(2k_x R_x-\varphi_R+2\pi/3)})\Big]\, ,\\
			\chi^{\rm HgPt}_{yz}(\epsilon,\vec{R}) ={} &- \mathcal{A}^2 e^{-2 i \pi/3} \Big[r r_- (e^{-i\varphi_R}+ e^{i (2k_x R_x-2k_y R_y+\varphi_R + 2\pi/3)}) + r r_+ (e^{-i\varphi_R}+ e^{i (-2k_x R_x+2k_y R_y+\varphi_R + 2\pi/3)})\notag \\ {} &
			+t t_- (e^{i\varphi_R}+ e^{-i (-2k_x R_x-2k_y R_y+\varphi_R - 2\pi/3)}) + t t_+ (e^{i\varphi_R}+ e^{-i (2k_x R_x+2k_y R_y+\varphi_R - 2\pi/3)})\notag \\ {} &
			-r t_- (e^{-i (2k_y R_y+\varphi_R)}+ e^{i (2k_x R_x+\varphi_R+ 2\pi/3)}) - r t_+ (e^{-i (-2k_y R_y+\varphi_R)}+ e^{i (-2k_x R_x+\varphi_R+ 2\pi/3)})\notag \\ {} &
			-t r_- (e^{i (2k_y R_y+\varphi_R)}+ e^{-i (-2k_x R_x+\varphi_R- 2\pi/3)}) - t r_+ (e^{i (-2k_y R_y+\varphi_R)}+ e^{-i (2k_x R_x+\varphi_R- 2\pi/3)})\Big]\, ,\\
			\chi^{\rm HgPt}_{zy}(\epsilon,\vec{R}) ={} & \mathcal{A}^2 e^{-2 i \pi/3} \Big[r r_- (e^{-i\varphi_R}+ e^{i (-2k_x R_x+2k_y R_y+\varphi_R + 2\pi/3)}) + r r_+ (e^{-i\varphi_R}+ e^{i (2k_x R_x-2k_y R_y+\varphi_R + 2\pi/3)})\notag \\ {} &
			+t t_- (e^{i\varphi_R}+ e^{-i (2k_x R_x+2k_y R_y+\varphi_R - 2\pi/3)}) + t t_+ (e^{i\varphi_R}+ e^{-i (-2k_x R_x-2k_y R_y+\varphi_R - 2\pi/3)})\notag \\ {} &
			-r t_- (e^{i (2k_y R_y-\varphi_R)}+ e^{i (-2k_x R_x+\varphi_R+ 2\pi/3)}) - r t_+ (e^{-i (2k_y R_y+\varphi_R)}+ e^{i (2k_x R_x+\varphi_R+ 2\pi/3)})\notag \\ {} &
			-t r_- (e^{i (-2k_y R_y+\varphi_R)}+ e^{-i (2k_x R_x+\varphi_R- 2\pi/3)}) - t r_+ (e^{i (2k_y R_y+\varphi_R)}+ e^{-i (-2k_x R_x+\varphi_R- 2\pi/3)})\Big]\, .
			\end{align}
		\end{subequations}    
		
		Many terms in the above susceptibilities cancel out for simple Dirac cones without distinct spin and valley degrees of freedom. However, with specific spin- and valley-dependent gaps $\Delta_{\nu,s}$, such simplifications are impossible, and one must deal with all terms.
		
		\begin{figure}[t]
			\centering
			\includegraphics[width=0.7\linewidth]{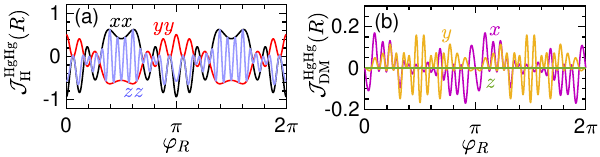}
			\caption{\textbf{RKKY interaction periodicity}. The cross-talk between spins and valleys in a pristine Kane-Mele 2D QSHI (monolayer Pt$_2$HgSe$_3$) leads to an expected beating pattern of RKKY oscillations. We set $R/a \approx 7$\, and $V = h = \Delta_{\rm d}$ = 0.} 
			\label{fap2}
		\end{figure}
		
		\section{RKKY interaction periodicity}\label{ap2}
		In this Appendix, we briefly show the RKKY interaction periodicity. The RKKY interaction exhibits an oscillatory behavior that depends on the distance between two magnetic impurities. As this distance changes, the interaction alternates. The period of this oscillation is governed by the Fermi wave vector, whether pristine or perturbed, which is crucial in determining the distance at which the interaction switches from FM/AFM to AFM/FM or from CW/CCW to CCW/CW. Given the assumed strong SOC in the employed Kane-Mele model, as in the case of monolayer Pt$_2$HgSe$_3$, and the cross-talk between spins and valleys, a beating pattern of oscillations is expected even in the pristine QSHI phase, as confirmed in Fig.~\ref{fap2}.
		
	}
	\twocolumngrid
	\bibliography{bib2.bib}
	
\end{document}